\documentclass{article}

\usepackage{glas}

\begin{document}

\def\Journal#1#2#3#4{{#1} {\bf #2}, #3 (#4)}
%
\def\NCA{\em Nuovo Cimento}
\def\NIM{\em Nucl. Instrum. Methods}
\def\NIMA{{\em Nucl. Instrum. Methods} A}
\def\NPB{{\em Nucl. Phys.} B}
\def\PLB{{\em Phys. Lett.}  B}
\def\PRL{\em Phys. Rev. Lett.}
\def\PRD{{\em Phys. Rev.} D}
\def\ZPC{{\em Z. Phys.} C}
\def\st{\scriptstyle}
\def\sst{\scriptscriptstyle}
\def\mco{\multicolumn}
\def\epp{\epsilon^{\prime}}
\def\vep{\varepsilon}
\def\ra{\to}
\def\ppg{\pi^+\pi^-\gamma}
\def\vp{{\bf p}}
\def\ko{K^0}
\def\kb{\bar{K^0}}
\def\al{\alpha}
\def\ab{\bar{\alpha}}
\def\be{\begin{equation}}
\def\ee{\end{equation}}
\def\bea{\begin{eqnarray}}
\def\eea{\end{eqnarray}}
\def\CPbar{\hbox{{\rm CP}\hskip-1.80em{/}}}
%

%
%


%

\begin{titlepage}{GLAS-PPE/2001--05}{\\[0mm]}
\title{\bf The ALEPH Search for the Standard Model Higgs Boson}
\author{John Kennedy\\
Department of Physics and Astronomy \\
         University of Glasgow \\
         Glasgow, UK}
\vspace{-0.5cm}
\collaboration{on behalf of the ALEPH collaboration}
\begin{abstract}
A search has been performed for the Standard Model Higgs boson in the data collected with the ALEPH detector in 2000. An excess of 3$\sigma$ above the background expectation is found. The observed excess is consistent with the production of the Higgs boson with a mass close to 114 GeV/$c^{2}$.
\end{abstract}

\end{titlepage}

\section{Introduction}\label{sec:intro}

This note presents results on the search for the Standard Model Higgs boson by the ALEPH collaboration in the year 2000. This search primarily looks for the production of Higgs bosons via the Higgsstrahlung process, although some search channels also posses a limited sensitivity to Higgs production via W and Z vector boson fusion. In total, 216.2$\mathrm{pb}^{-1}$ of data was collected at centre of mass energies ranging from 200 GeV to 209 GeV, with the majority collected around 205.1 GeV (72$\mathrm{pb}^{-1}$) and 206.7 GeV (107$\mathrm{pb}^{-1}$). 
\section{Event Selection}\label{sec:evt_sel}

The search analyses, described in detail elsewhere~\cite{pap189}$^{,}$\cite{pap2000}, are designed to detect specific ``channels'' or final states arising from the Higgsstrahlung production process: 

\begin{itemize}  
\item The four jet final state (h$q\bar{q}$)$^{\dagger}$
\item The missing energy final state (h$\nu\bar{\nu}$)$^{\dagger}$
\item The leptonic final state (h$\mbox{\ensuremath{{\ell}^{+}{\ell}^{-}}}$) 
\item The tau lepton final state (h$\tau\tau$ and h$\to \tau\tau$, Z   $\to q\bar{q}$) 
\end{itemize}

The Higgs boson search was conducted using both a Neural Network based stream (denoted ``NN'') and a cuts based stream (``cuts''). The two streams are formed from the above four final state analyses where final states marked with a $\dagger$ have alternate analyses based on NN's and cuts and the searches are identical in both streams for the leptonic and tau final states. The analysis selection criteria were fixed before the data taking period began thus ensuring that their application to the collected data was unbiased.

\section{Results and Statistical Interpretation}\label{sec:res_Int}

The number of observed and expected events for each analysis channel and the total combined NN and cuts streams are given in table \ref{tab:evt_numbers}.

\begin{table} [h!!!] 
\caption{The number of expected signal and background events for each analysis channel with the expected significance and number of observed candidates.}
\label{tab:evt_numbers}
\begin{center}
\footnotesize
\begin{tabular}{|c|c|c|c|c|} \hline
Search & Expected & Expected & Events & Expected \\
channel &background & Signal & Observed & Significance($\sigma$) \\ \hline
4-jet (NN) &  46.9 & 4.5 &52  & 1.6 \\
4-jet (Cut) & 23.7 & 2.9 &31  & 1.3 \\
h$\nu\nu$ (NN) &37.5 &1.4 &38 & 0.8 \\
h$\nu\nu$ (Cut) &19.7 &1.3 &20 & 0.7 \\
h$\ell\ell$ &30.6 &0.7 &29  &0.8  \\
$\tau\tau q\bar{q}$ &13.6 &0.4 &15 &0.4  \\ \hline
Tot (NN) &128.7 &7.0 &134 &2.1  \\ \hline
Tot (Cut) &87.6 &5.3 &95 & 1.8 \\ \hline
\end{tabular}
\end{center}
\end{table}

\begin{figure}[t]
\begin{center}
\begin{picture}(400,200)
\put(-10,-20){\epsfxsize=15pc\epsfbox{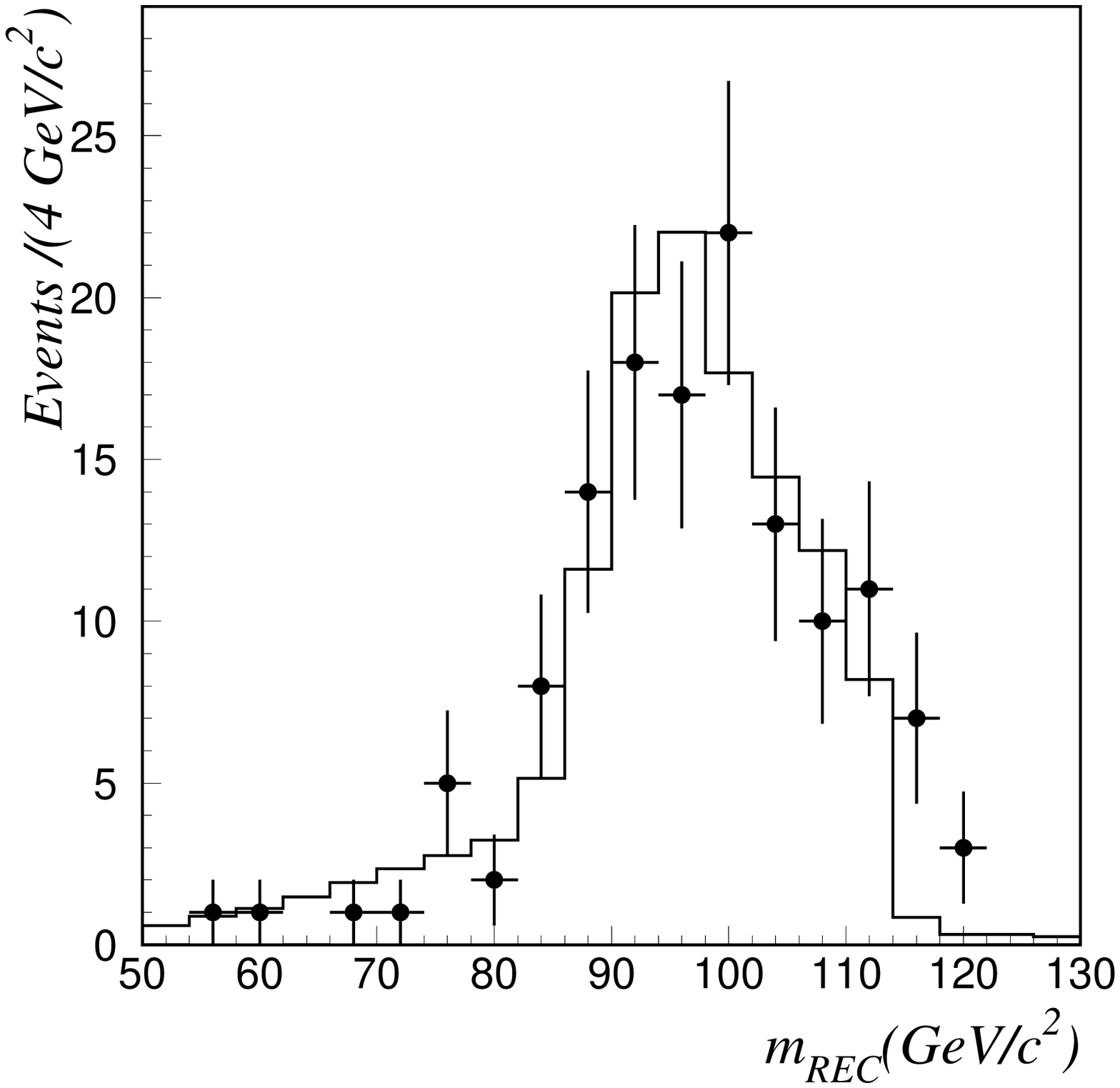}}
\put(160,-20){\epsfxsize=15pc\epsfbox{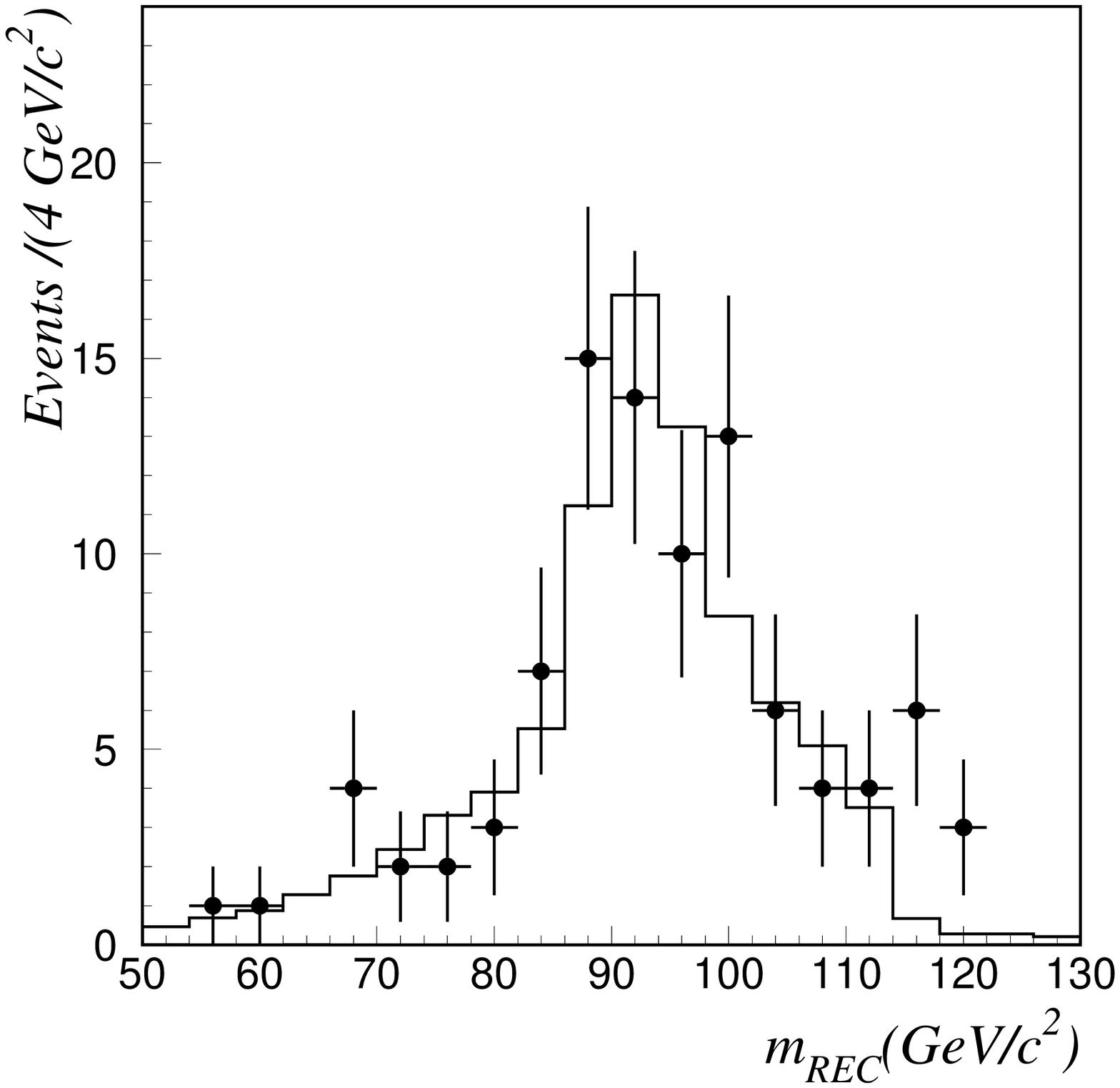}}
\put(20,130){(a)}
\put(190,130){(b)}
\end{picture}
\end{center}
\caption{Distributions of the reconstructed Higgs boson mass for the data collected in 2000(points) and the expected background(histogram) for the NN(a) and cuts(b) streams.}
\label{fig:hmass}
\end{figure}

Figures \ref{fig:hmass}a and \ref{fig:hmass}b show the distributions of the reconstructed Higgs boson mass($m_{REC}=m_{12}+m_{34}-91.2$) for the data and expected background in the NN and cuts based streams respectively. Both figures show good agreement between data (points) and the expected background (histogram) at low reconstructed Higgs masses while they both display a clear excess of data candidates at large reconstructed Higgs masses.

\subsection{Confidence Level Results}

The mass is not the only information which may be used to distinguish a Higgs boson signal from SM background. Further information (eg, b-tag, NN$_{output}$) is taken into account in the construction of the likelihood ratio $Q = L_{s+b}/L_{b}$, where $L_{b}$ is the likelihood for the background-only hypothesis and $L_{s+b}$ is the likelihood for the signal+background hypothesis for a given Higgs boson mass. The likelihood ratio measures the compatibility of the experiment with a given signal mass hypothesis and is defined as:

   \begin{equation}
      Q = \frac {L_{s+b}}{L_{b}}=\frac{exp^{-(s+b)}}{exp^{-b}}
      \prod_{i=1}^{n_{obs}}\frac{sf_{s}(X_{i})+bf_{b}(X_{i})}{bf_{b}(X_{i})}
   \end{equation}

where $s$ and $b$ are the total number of signal and background events expected. The functions $f_{s}$ and $f_{b}$ are the probability densities that a signal or background event will be found in a given final state identified by a set of discriminating variables. By removing these functions, the likelihood ratio is returned to the ratio of the Poisson probabilities to observe $n_{obs}$ events for the signal+background and background-only hypotheses.

\begin{figure}[t]
\begin{center}
\begin{picture}(400,200)
\put(-10,-20){\epsfxsize=15pc\epsfbox{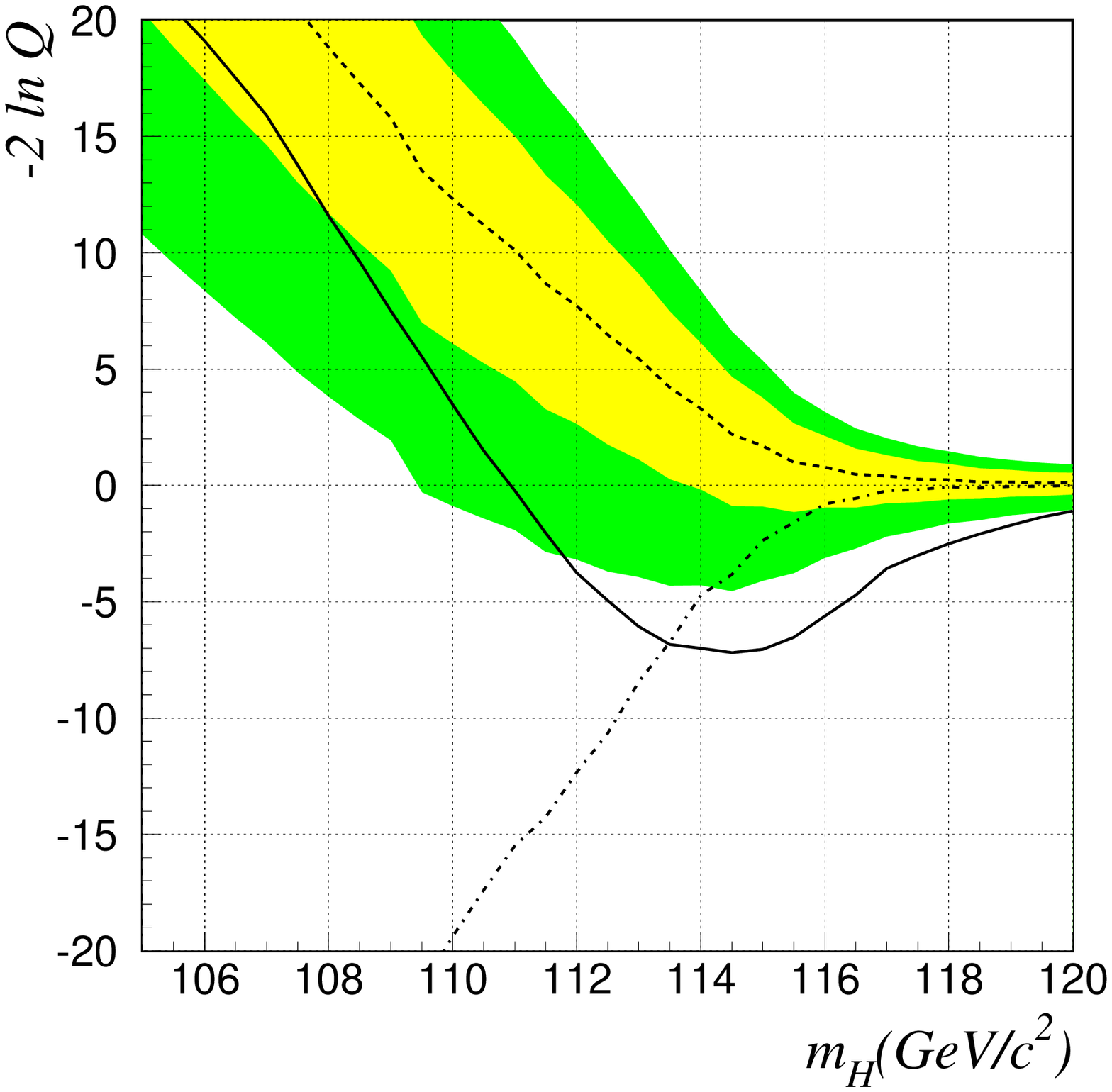}}
\put(160,-20){\epsfxsize=15pc\epsfbox{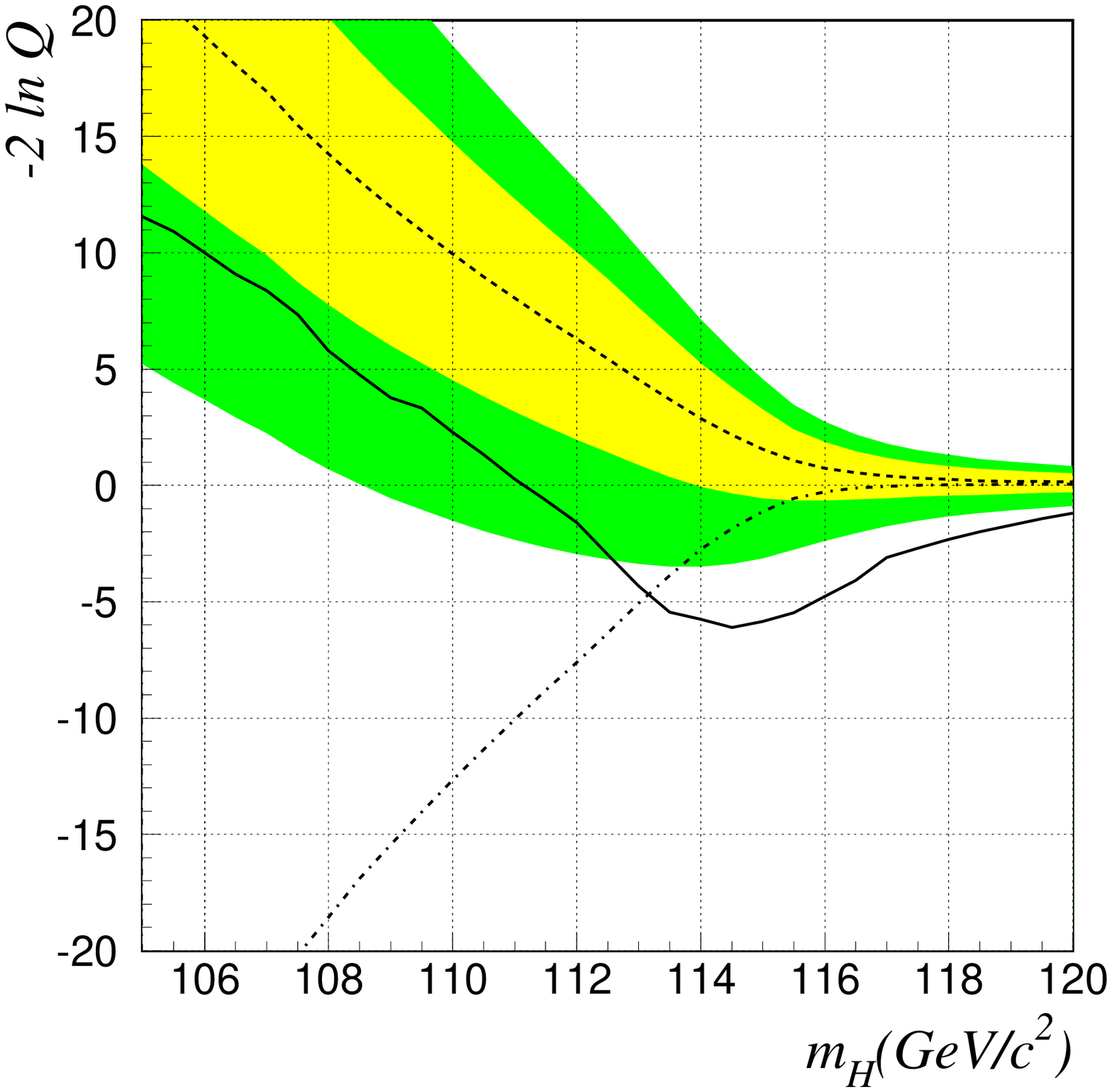}}
\put(135,130){(a)}
\put(305,130){(b)}
\end{picture}
\end{center}
\caption{The log-likelihood estimator -2ln$Q$ for the NN(a) and cuts(b) streams as a function of the hypothesised Higgs boson mass. Observed(solid), background only expectation(dashed) and signal(dot-dash) are shown with light and dark bands around the background expectation representing the one and two sigma bands respectively.}
\label{fig:lnq}
\end{figure}

The observed and expected distributions of the likelihood ratio, expressed as -2ln$Q$, are shown in figures \ref{fig:lnq}a and \ref{fig:lnq}b for the NN and cuts streams respectively. 
 
The compatibility of an experiment with a given hypothesis is derived by calculating the probability of obtaining a likelihood ratio smaller than that observed. This probability is referred to as the confidence level($CL$). When observed results are tested against a background-only hypothesis we use the quantity $CL_{b}$. In the case of the expected SM background $CL_{b}$ has a median value at 0.5 while the observation of a signal is expected to produce an excess relative to the background and the value of $CL_{b}$ (1-$CL_{b}$) is expected to rise (drop).

\begin{figure}[t]
\begin{center}
\begin{picture}(400,200)
\put(-10,-20){\epsfxsize=15pc\epsfbox{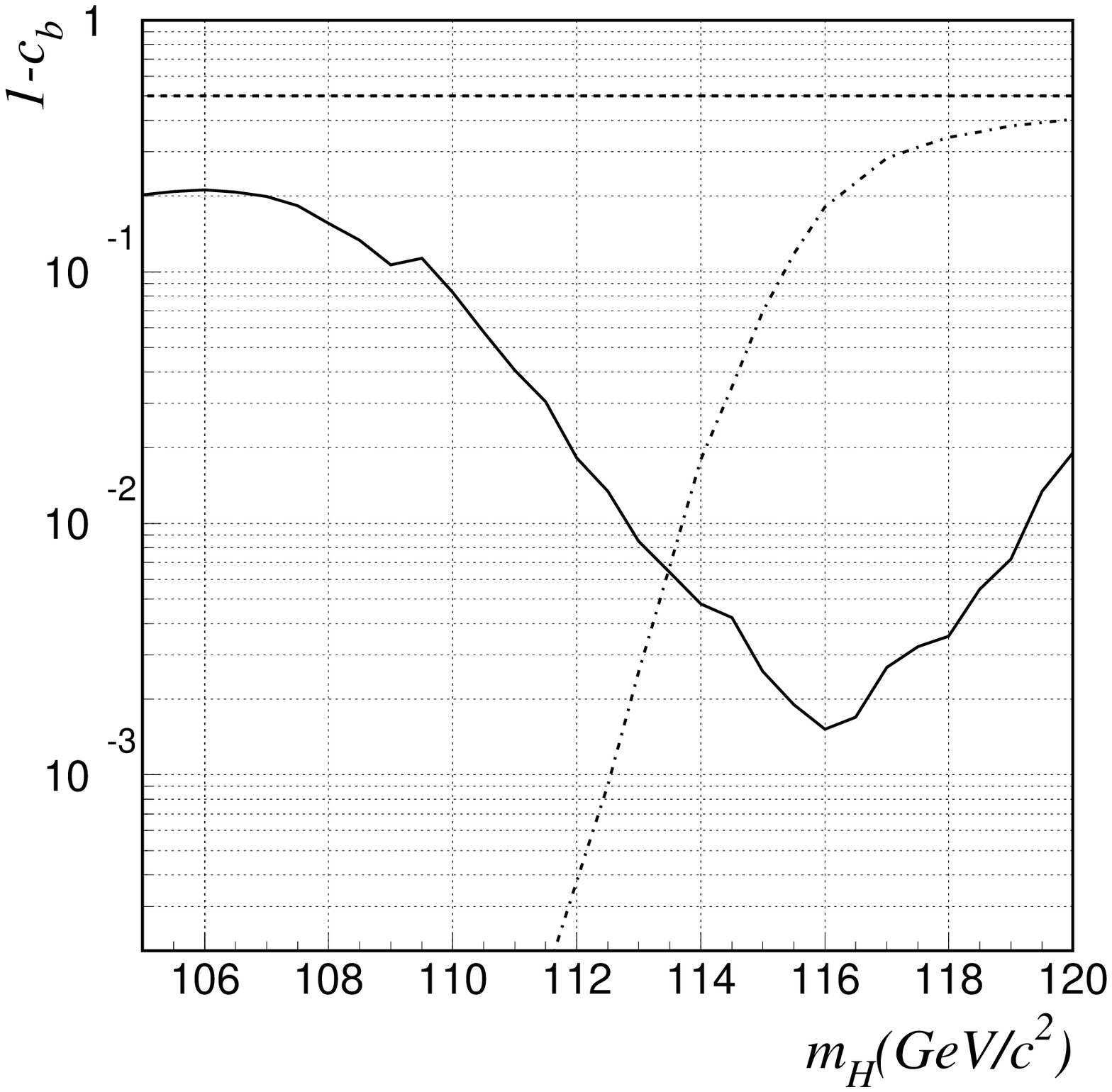}}
\put(160,-20){\epsfxsize=15pc\epsfbox{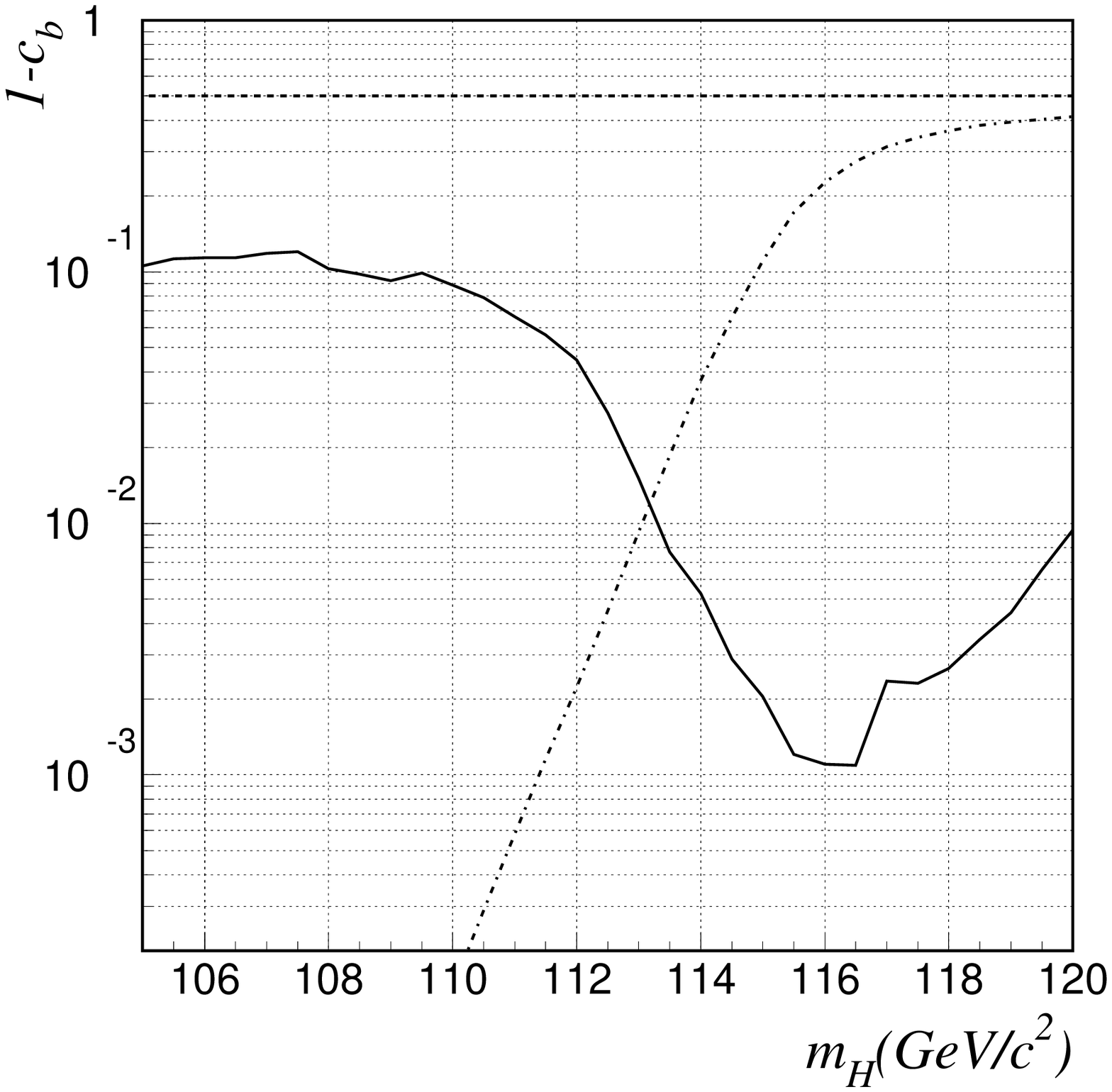}}
\put(20,130){(a)}
\put(190,130){(b)}
\end{picture}
\end{center}
\caption{Distributions of the $CL_{b}$ curves for the Observed(solid), expected background(dashed) and signal+background(dot-dash) for the NN(a) and cuts(b) streams.}
\label{fig:cb}
\end{figure}

Figures \ref{fig:cb}a and \ref{fig:cb}b show the observed and expected distributions for 1-$CL_{b}$
as a function of the hypothesised Higgs boson mass. A large deviation from the expected background value of 0.5 can be seen in both the NN and cuts based streams. This deviation is consistent with an excess of events over the background only hypothesis and is maximal for a Higgs boson mass of $\approx$ 116 GeV$/c^{2}$. The probability of deviations of this magnitude are 1.5 x $10^{-3}$ and  1.1 x $10^{-3}$ corresponding to significances of 3.0$\sigma$ and 3.1$\sigma$ \footnote{The ALEPH significances are calculated using single sided Gaussian distributions} for the NN and cuts streams respectively.

\subsection{Analysis of `High Impact' Candidates}

The `Quality' of individual candidates may be assessed by two independent methods. These methods allows us to gain information about the individual candidates and also assess the origin and stability of the observed excess.

\begin{figure}[t]
\begin{center}
\begin{picture}(400,200)
\put(-10,-20){\epsfxsize=15pc\epsfbox{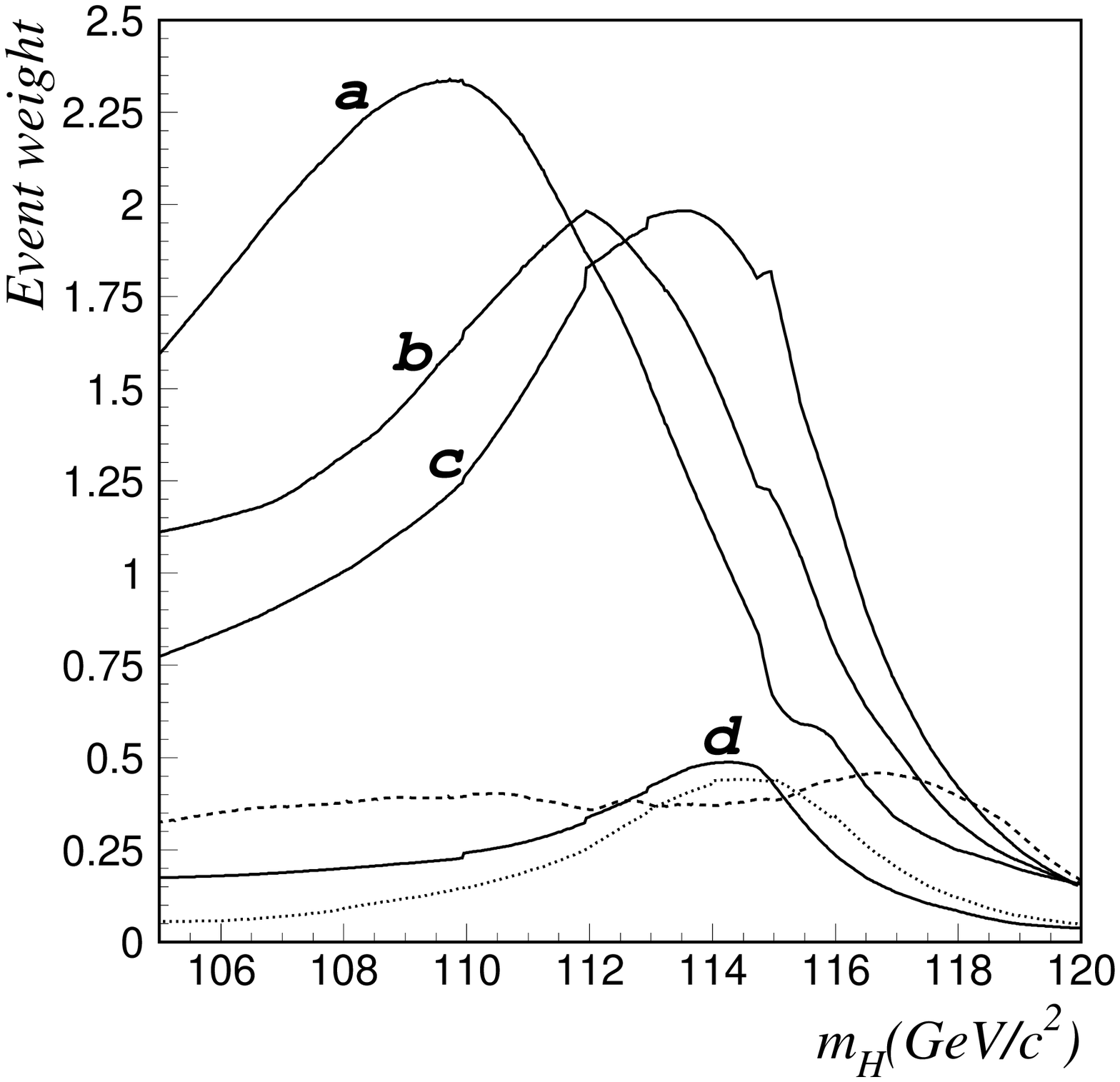}}
\put(160,-20){\epsfxsize=15pc\epsfbox{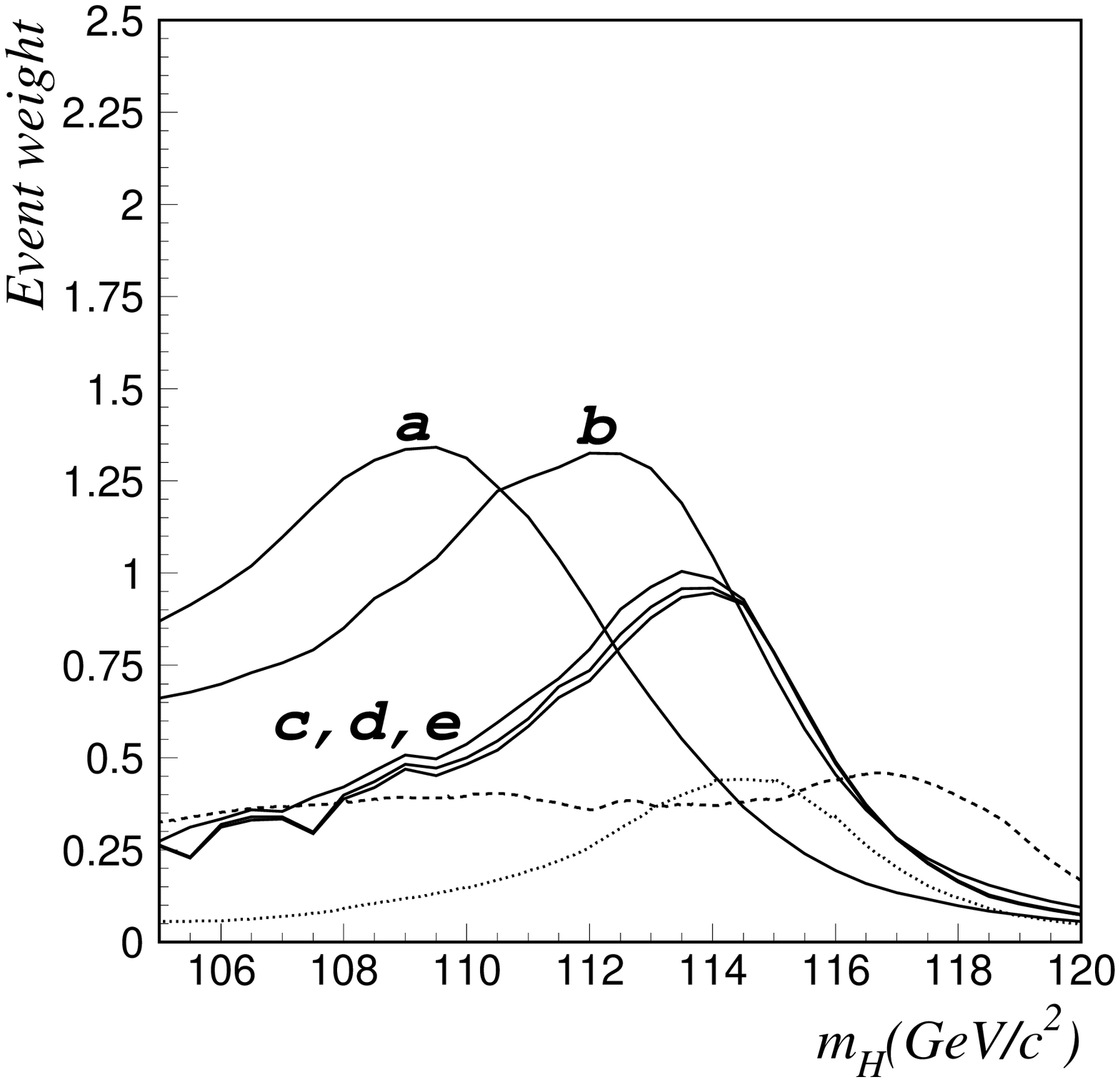}}
\put(135,130){(a)}
\put(305,130){(b)}
\end{picture}
\end{center}
\caption{Event weights for the individual candidates as a function of the hypothesised Higgs boson mass for the NN(a) and cuts(b) based streams. Contributing candidates are from the four jet(solid), leptonic(dashed) and tau(dotted) channels.}
\label{fig:weight}
\end{figure}

The impact of each candidate may be determined by calculating the contribution of the candidate to -2ln$Q$. In the calculation of -2ln$Q$, individual candidates contribute as a sum of event weights, ln(1+$\frac{sf_{s}}{bf_{b}}$), which may then be analysed. Figures \ref{fig:weight}a and \ref{fig:weight}b show the event weights, as a function of hypothesised Higgs boson mass, for all candidates with weights larger than 0.4 at a mass of 114 GeV$/c^{2}$ for the NN and cuts analyses respectively. The details of the highlighted `high impact' candidates, a-e, all of which occur in the four jet final state, are found in table \ref{tab:candidates}.

\begin{table}[t]
\caption{Details of the `High Impact' candidates with weights above 0.4 at a Higgs boson mass of 114 GeV/$c^{2}$.}
\label{tab:candidates}
\begin{center}
\footnotesize
\begin{tabular}{|c|c|c|c|c|c|c|c|c|}
\hline
Event & Higgs Mass & $m_{12}$ & $m_{34}$ & b-tag & b-tag & b-tag & b-tag & 4-Jet \\
 & (GeV$/c^{2}$) & (GeV$/c^{2}$) & (GeV$/c^{2}$) & Jet 1 & Jet 2 & Jet 3 & Jet 4 & NN \\ \hline 
\em{a}& 110.0 & 96.3 & 104.9 & 0.999 & 0.836 & 0.999 & 0.214 & 0.999 \\
\em{b}& 112.9 & 94.9 & 109.2 & 0.994 & 0.776 & 0.993 & 0.999 & 0.997 \\ 
\em{c}& 114.3 & 101.3 & 104.2 & 0.136 & 0.012 & 0.999 & 0.999 & 0.996 \\
\em{d}& 114.5 & 78.8 & 126.9 & 0.238 & 0..52 & 0.998 & 0.948 & 0.935 \\ 
\em{e}& 114.6 & 79.7 & 126.1 & 0.008 & 0.293 & 0.895 & 0.998 & 0.820 \\ \hline \end{tabular}
\end{center}
\end{table}

\begin{figure}[t]
\begin{center}
\begin{picture}(400,200)
\put(-10,-20){\epsfxsize=15pc\epsfbox{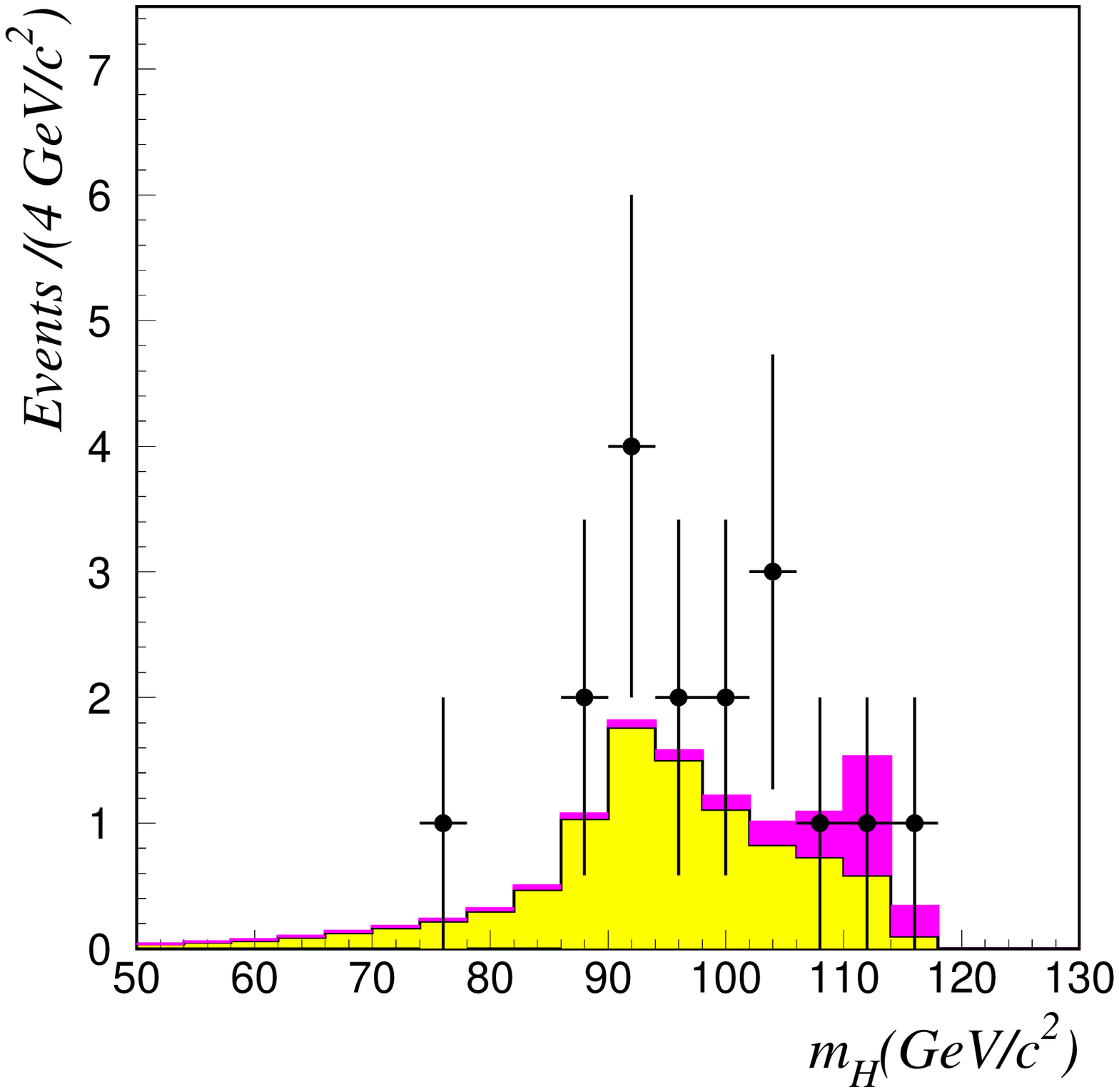}}
\put(160,-20){\epsfxsize=15pc\epsfbox{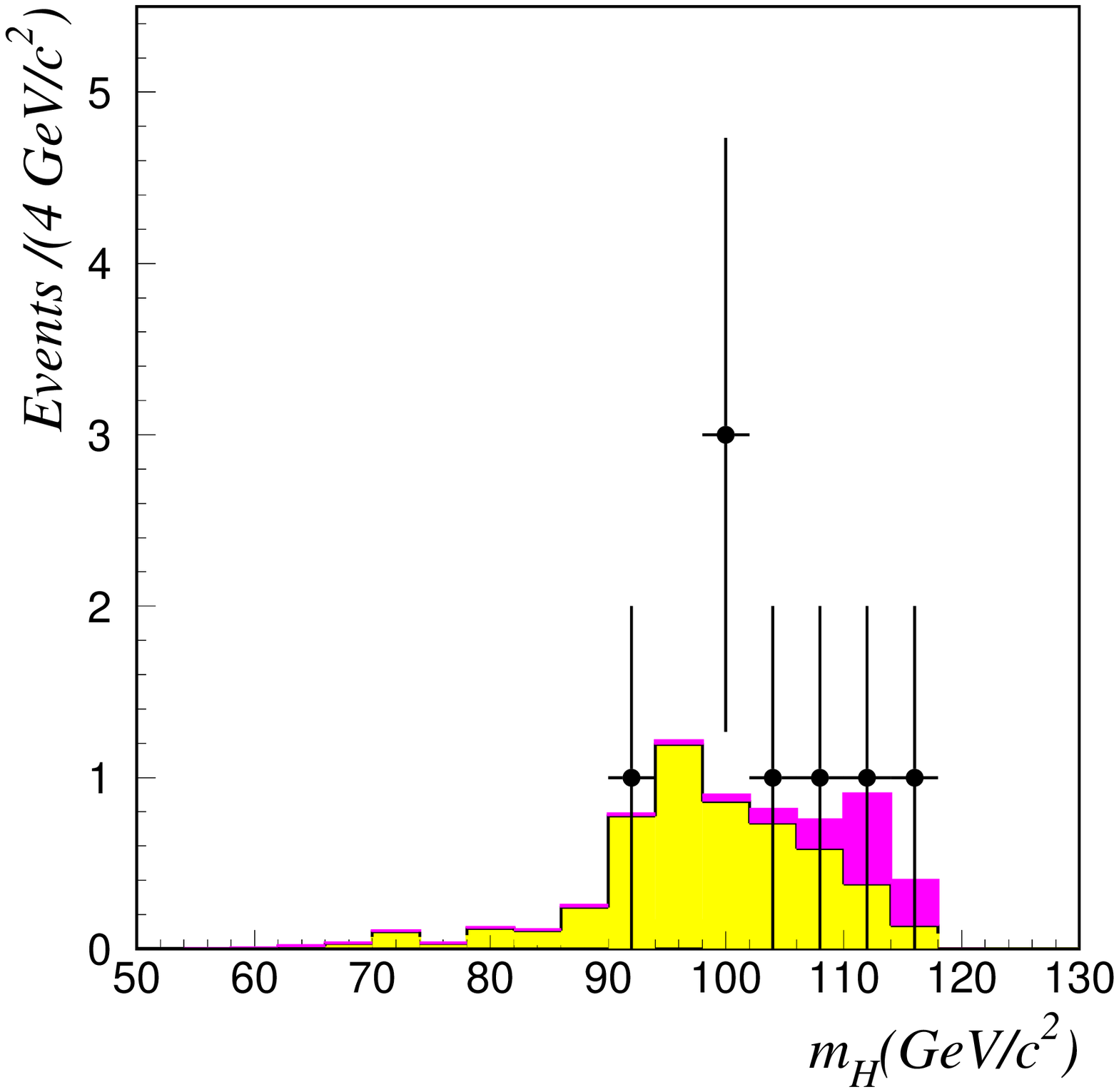}}
\put(20,130){(a)}
\put(190,130){(b)}
\end{picture}
\end{center}
\caption{High `Purity'  $(s/b)_{109}$ reconstructed Higgs boson mass distributions for the NN(a) and cuts(b) streams. Data(points with errors), expected background(light histogram) and expected signal with $m_{h}=114$GeV$/c^{2}$(dark histogram).}
\label{fig:purity}
\end{figure}

Alternatively the `Quality' of the candidates may be assessed by considering their `purity'. Here we define purity to be the ratio of expected number of signal to background events with a reconstructed Higgs boson mass greater than 109 GeV$/c^{2}$, from here on denoted as $(s/b)_{109}$. Tightening the selection criteria of the analyses enables the purity to be increased and allows us to gain information about the candidate `Quality' and also the stability of the observed excess. High purity values $(s/b)_{109}$=1.5 are achieved by tightening the NN cut on the four jet selection in the NN stream and tightening b-tag/Z-mass constraints on the four jets selection in the cuts stream. Figures \ref{fig:purity}a and \ref{fig:purity}b show the high purity distributions of the reconstructed Higgs boson mass for the NN and cuts streams respectively.

A large overlap between the high `weight' and high `purity' candidates adds confidence to the observed result and the stability of the excess.

\section{Conclusion}\label{sec:conc}

The data collected with the ALEPH detector in 2000 have been analysed to search for the Standard Model Higgs boson. Both NN and Cuts based analysis streams show a 3$\sigma$ excess beyond the Standard Model background expectation. The observed result is consistent with the production of a Higgs boson with a mass near 114 GeV$/c^{2}$.

\end{document}